\documentstyle[11pt,twoside,psfig,jltp]{article}
\begin{document}
\title{Screening, nonadiabaticity, and quantized acoustoelectric current }
\author{Michael Pustilnik$^{a,b}$, Karsten Flensberg$^{b}$, and Qian Niu$^{c}$}
\address{$^{a}$ Danish Institute of Fundamental Metrology, \\
Anker Engelunds Vej 1, Building 307, DK-2800 Lyngby, Denmark \\
$^{b}$ \O rsted Laboratory, Niels Bohr Institute, \\
Universitetsparken 5, DK-2100 Copenhagen, Denmark \\
$^{c}$ Department of Physics, The University of Texas at Austin, \\
Austin, TX 78712, USA }

\begin{abstract}
Quantized single-electron transport driven by surface acoustic waves (SAW)
through a pinched-off narrow constriction is studied theoretically.
Long-range Coulomb interaction causes the tunneling coupling between the
two-dimensional electron gas (2DEG) and the moving minimum of the
SAW-induced potential to decay rapidly with time. The energy scale $\hbar
/\tau $, associated with the characteristic time of this decay $\tau $,
controls both the width of the transition regions between the plateaus and
the slope of the plateaus. This sets a limit for the accuracy of the
quantization of acoustoelectric current at low temperature.

\end{abstract}

\maketitle

\runninghead{M. Pustilnik, K. Flensberg, and Q. Niu}
{Screening, nonadiabaticity, and quantized acoustoelectric current}

\vspace{0.3in}

Recent experiments \cite{exp}$^,$\cite
{openchannel} demonstrated that surface acoustic waves (SAW) induce
charge transport through a narrow constriction, formed in GaAs
heterostructure by a split-gate depletion technique. In the pinch-off 
regime \cite{exp} the acoustoelectric
current, as function of the potential at the gate electrode, exhibits
plateaus, where 
\begin{equation}
I_{ae}=N_{0}ef.  \label{quantization}
\end{equation}
Here $f$ is SAW frequency, $e$ is the electron charge, and $N_{0}$ is an
integer. The plateaus were demonstrated to be stable over the range of
temperature, SAW power, and applied source-drain voltage. The remarkable
accuracy of the quantization (of the order of $10^{-5}$), and high frequency
of operation \cite{frequency} immediately suggest possible metrological
applications of the effect as a tool for maintaining an independent current
standard \cite{metrology} . It is therefore important to the
underlying physics of the effect, and, therefore, possible sources of errors.

Qualitatively, the effect is explained by a simple picture of moving quantum
dots \cite{exp}. Electrons, trapped in the moving minima ('dots') of
SAW-induced potential, are dragged through the potential barrier. The strong
Coulomb repulsion prevents excess occupation of the dot. Increase of SAW
power deepens the dots, more states in the dot become available for
electrons to occupy, and new plateaus appear. The slope of the potential
barrier can be lowered by changing the gate voltage, which has a similar
effect.

\begin{figure}[tbp]
\centerline{\vspace{2mm}}
\centerline{\psfig{file=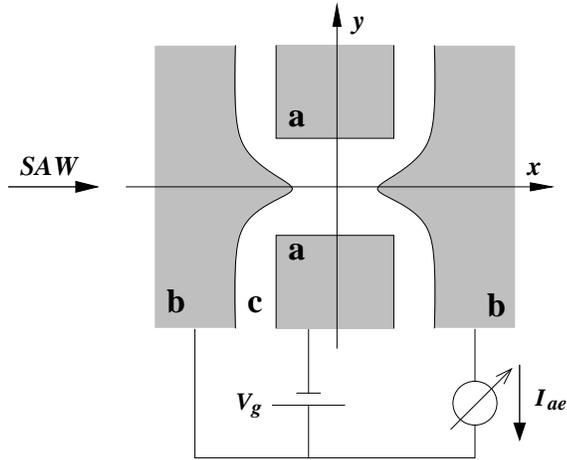,width=7.5cm}}
\centerline{\vspace{-0.25mm}}
\caption{SAW device: (a) split gates; (b) regions, occupied by
two-dimensional electron gas; (c) depleted region. The arrow indicates
direction of SAW propagation.\protect\vspace{1mm}}
\end{figure}

In principle, quantization of the acoustoelectric current in one-dimensional
channels is expected on the quite general theoretical ground \cite{open}.
However, the effect was not observed in the open channel regime \cite
{openchannel}. The reason is that the mechanism of quantization \cite{open}
requires that the DC conductance for each instantaneous configuration of
SAW-induced potential is zero. This can be achieved, if the channel is long
(much longer than SAW wavelength $\lambda _{s}$), and the Fermi level lies
in the gap of the spectrum. These conditions are difficult to realize
experimentally \cite{openchannel}. In the alternative approach, put forward
in \cite{exp}, the channel is biased beyond the pinch-off (see Fig. 1). This
ensures that the DC conductance is zero. However, as shown below, the rapid
change of SAW-induced potential near the entrance of the channel may lead to
significant non-adiabatic corrections to (\ref{quantization}).

Since the Fermi velocity $v_{s}$ is much larger than the sound velocity $%
v_{s}$, the 2DEG is able to follow the changing in time SAW-induced
potential. Therefore, the effective potential $\varphi \left( x,y\right) $,
as seen by the electrons in the system, can be determined from a
self-consistent solution of the instantaneous electrostatic problem,
combined with the Thomas-Fermi-type relation between the density of the
electron gas and $\varphi $. This problem is still extremely complex.
However, important properties of the solution can be understood as follows.
\begin{figure}[tbp]
\centerline{\vspace{2mm}}
\centerline{\psfig{file=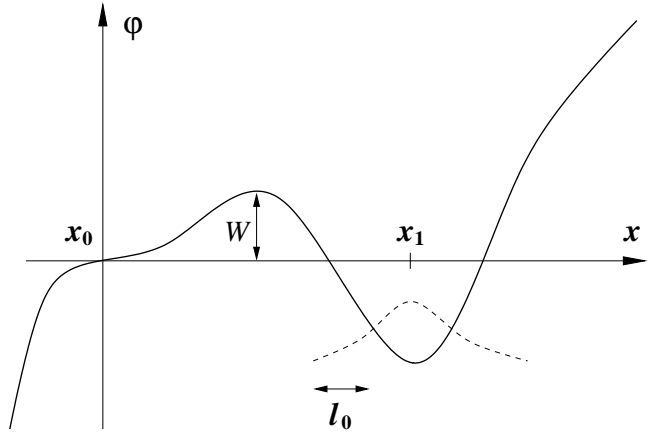,width=8.5cm}}
\centerline{\vspace{-0.25mm}}
\caption{The instantaneous electrostatic potential $\protect\varphi \left(
x,y\right) $ along the line $y=0$. 2DEG is confined to the region $x<x_{0}$
(so that the line $\protect\varphi =0$ coincides with the Fermi level). The
location $x_{1}$ of the SAW-induced potential minimum moves to the right
with the sound velocity $v_{s}$. The dashed line shows the wave function of
the trapped electron.\vspace{1mm}}
\end{figure}
The crucial observation is that, since the screening length in 2DEG ($\sim
10\;{\rm nm}$) is small compared to SAW wavelength ($\lambda _{s}\sim 1\;\mu 
{\rm m}$), the SAW-induced potential is screened almost completely in the
regions, occupied by 2DEG (see Fig. 1,2). On the other hand, the screening in
the depleted region is lacking. The SAW induce a potential minimum (the
'dot'), located near some point $x_{1}$ in the depleted region (see Fig. 2).
As the system evolves ($\varphi $ depends parametrically on time), $x_{1}$
moves to the right with the sound velocity $v_{s}$. At the same time, the
position of the edge of 2DEG, $x_{0}$, is expected to oscillate with SAW
frequency $f$. However, the amplitude of these oscillations is small (since
the amplitude of the SAW-induced potential is definitely small, compared to $%
V_{g}$). Therefore, the width of the potential barrier, that separates 2DEG
and the dot, grows linearly with time. As a result, the tunneling amplitude
decays approximately exponentially with the characteristic time $\tau \sim
l_{0}/v_{s}$, where $l_{0}$ is the distance over which the localized wave
function extends under the potential barrier (see Fig. 2). Since, evidently, $%
l_{0}\ll \lambda _{s}$, the inequality 
\begin{equation}
f\tau \ll 1  \label{separation}
\end{equation}
holds. Due to this inequality, the time-dependence of all other parameters
of the system can be neglected. For example, the change $\delta \varepsilon
_{0}$ of the energy of the localized level $\varepsilon _{0}$ during the
time $\tau $ can be estimated as  
\[
\delta \varepsilon _{0}=\varepsilon _{0}\left( t+\tau \right) -\varepsilon
_{0}\left( t\right) \sim \left( d\varepsilon _{0}/dt\right) \tau \sim
\varepsilon _{0} f\tau \ll \varepsilon _{0},
\]
and is negligibly small. 

The rapid decay of the tunneling coupling results in nonadiabatic
corrections to (\ref{quantization}). These corrections are controlled by the
energy scale $\hbar /\tau $. 

To obtain an order of magnitude estimate of $\tau $, we expand the potential 
$V\left( x\right) $, seen by the electrons, near the minimum $x_{1}$ (see
Fig.2), we obtain $V\left( x\right) \approx A_{s}q_{s}^{2}\left(
x-x_{1}\right) ^{2}/2$, where $q_{s}=2\pi /\lambda _{s}$. The amplitude $%
A_{s}$ is related to the single-particle level spacing in the dot $\Delta $
via 
\[
A_{s}q_{s}^{2}=m^{\ast }\left( \Delta /\hbar \right) ^{2},
\]
where $m^{\ast }$ is an effective mass, to the 'size' $r$ of the wave
function of the localized electron via 
\[
A_{s}q_{s}^{2}r^{2}\sim \Delta ,
\]
and to the charging energy $E_{c}$ via 
\[
E_{c}\sim e^{2}/\epsilon r.
\]
$l_{0}$ can be estimated from the relation 
\[
\frac{\left( \hbar /l_{0}\right) ^{2}}{2m^{\ast }}\sim W+\varepsilon _{0},
\]
where $W$ is barrier's height, and $\varepsilon _{0}$ is the energy of the
localized level. Assuming that $W+\varepsilon _{0}\sim A_{s}$, this gives us
4 equations for 5 unknown quantities. Turning to the experiments, we note
that the quantization disappears above the activation temperature $T^{\ast
}\sim 10\,{\rm K}$, which we identify as the energy it takes to overcome the
Coulomb barrier. Substituting $E_{c}\sim k_{B}T^{\ast }\sim 1\,{\rm meV}$,
together with $m^{\ast }=0.07\,m_{0}$ ($m_{0}$ is a free electron mass), $%
\epsilon \approx 13$, $\lambda _{s}=10^{-4}\,{\rm cm}$ to the equations
above, we obtain $A_{s}\sim 0.5\,{\rm meV}$, $r\sim 100\,{\rm nm}$, $\Delta
\sim 0.1\,{\rm meV}$. Using $v_{s}=3\times 10^{5}\,{\rm cm/s}$, we finally
arrive at $\tau \sim 10\;{\rm ps}$. Since the corresponding energy scale $%
\hbar /\tau \sim 0.1\,{\rm meV}$ is finite (although small), the
nonadiabaticity may affect significantly the accuracy of the quantization at
low temperature. 

This can be understood from the following model \cite{orthodox} , in which the
Coulomb repulsion in the dot is treated by introducing a single parameter -
the charging energy $E_{c}$: 
\begin{eqnarray}
{\cal H} &=&\sum_{k\sigma }\xi _{k}c_{k\sigma }^{\dagger }c_{k\sigma
}+\sum_{n\sigma }E_{n}d_{n\sigma }^{\dagger }d_{n\sigma }+E_{c}\left( N-%
{\cal N}_{g}\right) ^{2}  \label{model} \\
&&+V\left( t\right) \sum_{kn\sigma }\left( c_{k\sigma }^{\dagger }d_{n\sigma
}+{\rm H.c.}\right) .  \nonumber
\end{eqnarray}
Here $\xi _{k}$ and $E_{n}$ are the single-particle energy levels in the
lead and in the dot correspondingly, $E_{c}$ is charging energy, $%
N=\sum_{n\sigma }d_{n\sigma }^{\dagger }d_{n\sigma }$ is number of electrons
in the dot, and ${\cal N}_{g}$ is a linear function of the gate voltage $%
V_{g}$. In writing the Hamiltonian (\ref{model}), we have assumed that while
the dot is in the vicinity of the left electrode, the tunneling to the right
is negligible. We have also assumed that the electron gas is in
thermodynamic equilibrium at all times (by virtue of the inequality $%
v_{s}\ll v_{F}$). According to the discussion above, we have taken into
account the time-dependence of the tunneling coupling only, 
\begin{equation}
V\left( t\right) =V_{0}e^{-t/\tau }.  \label{V}
\end{equation}
Our task is to calculate the occupation of the dot $N_{0}=\left\langle
N\right\rangle _{t=\infty }$ at $t\rightarrow \infty $, given that the system is in 
thermodynamic equilibrium at $t=-\infty $. The
acoustoelectric current is related to $N_{0}$ via (\ref{quantization}). The
model (\ref{model}) answers the minimal requirements: it produces the
correct adiabatic limit ($\tau \rightarrow \infty $), reproducing the
staircase-like dependence of $I_{ae}$ on the gate voltage, while allowing
explicitly for a non-equilibrium occupation of the dot.

The effect of the time-dependence of the tunnelling coupling depends on how
close the system is to the Coulomb blockade degeneracy points (half-integer $%
{\cal N}_{g}$). Away from these points, when the inequality 
\begin{equation}
2E_{c}\left| {\cal N}_{g}-n_{0}-1/2\right| \gg \max \left\{ T,1/\tau
\right\}   \label{adiabaticity}
\end{equation}
is satisfied, the time-dependence is too slow to cause the transitions
between the charge states of the dot, and in this respect the adiabatic
approximation is justified. Here $n_{0}$ is the integer part of ${\cal N}_{g}
$ and we use $\hbar =k_{B}=1$ for the remaining part of the text. The
occupation of the dot is given by the standard Coulomb blockade expression,
with the temperature replaced by the effective temperature $T_{eff}\sim \max
\left\{ T,1/\tau \right\} $, and $N_{0}$ is integer, apart from the
exponentially small corrections. Close to the degeneracy points, when the
inequality (\ref{adiabaticity}) breaks down, the time-dependence of the
tunneling coupling is fast enough to induce the transitions between the
different charge states of the dot. This immediately implies that the width
of the transition region is given by $T_{eff}$.

To illustrate these conclusions, we consider the limit when $1/\tau \ll \Delta
\ll E_{c}$, and neglect the spin of the electrons. This might seem to be an
oversimplification, but the result of the calculations for the more general
case is qualitatively the same. If
we also restrict our attention to the region, which includes only one
transition between the plateaus, $n_{0}<{\cal N}_{g}<n_{0}+1$ is satisfied,
we can take into account only two charge states of the dot, that with $%
N=n_{0}$, and that with $N=n_{0}+1$ electrons. Projected onto these states,
the Hamiltonian is written in terms of the fermion operator $d=\left|
n_{0}\right\rangle \left\langle n_{0}+1\right| $:
\begin{equation}
{\cal H}=\sum_{k}\xi _{k}c_{k}^{\dagger }c_{k}+E_{0}d^{\dagger }d+V\left(
t\right) \sum_{k}\left( c_{k}^{\dagger }d+{\rm H.c.}\right) ,
\label{spinless}
\end{equation}
where 
\[
E_{0}=2E_{c}\left( 1/2+n_{0}-{\cal N}_{g}\right) .
\]
The simplified model (\ref{spinless}) has an advantage of being
exactly solvable for arbitrary $V(t)$. The solution can be
obtained in various ways. One possibility is to use the equation-of-motion
technique \cite{JWM} to derive a first order differential equation for 
\[
\left\langle N\left( t\right) \right\rangle =n_{0}+\left\langle d^{\dagger
}(t)d(t)\right\rangle ,
\]
which results in the exact expression for $N_{0}$ in the form of the
functional of the time-dependent level width $\Gamma \left( t\right) =2\pi
\nu V^{2}\left( t\right) $, where $\nu $ is density of states at the Fermi
level. With $V(t)$ given by (\ref{V}), the result is written in the compact
form 
\begin{equation}
N_{0}-n_{0}=\frac{\tau _{0}}{2\pi }\int_{-\infty }^{\infty }d\omega \frac{%
n_{F}(\omega )}{\cosh \left[ \pi \left( \omega -E_{0}\right) \tau _{0}\right]
},  \label{result}
\end{equation}
where $n_{F}$ is the Fermi function, and $\tau _{0}=\pi \tau /2$. At $T=0$ it
reduces to 
\[
N_{0}-n_{0}=\frac{2}{\pi }\tan ^{-1}\left( e^{-E_{0}\tau _{0}}\right) 
\]
The result (\ref{result}) is described very well by the Fermi function 
\begin{equation}
N_{0}-n_{0}\approx \left( e^{E_{0}/T_{eff}}+1\right) ^{-1},\;T_{eff}=\sqrt{%
T^{2}+\left( c/\tau _{0}\right) ^{2}}.  \label{Fermi}
\end{equation}
We found that $c=0.88$ gives the best numerical fit.

Close to the middle of the plateau (${\cal N}_{g}\rightarrow n_{0}$), Eqs. (%
\ref{quantization}) and (\ref{Fermi}) give the following expression for the
plateau's slope: 
\begin{equation}
S=\frac{1}{I_{0}}\left( \frac{dI_{ae}}{d{\cal N}_{g}}\right) _{{\cal N}%
_{g}\rightarrow n_{0}}\approx \left( 2E_{c}/T_{eff}\right)
e^{-E_{c}/T_{eff}},  \label{slope}
\end{equation}
where $I_{0}=n_{0}ef$ corresponds to the ideal quantization. 
Strictly speaking, the two-sate approximation is not
sufficient to obtain the correct value of $S$ precisely at the middle of the
plateau: the state with $N=n_{0}-1$ makes exactly the same contribution, as
that with $N=n_{0}+1$. This complication, however, does not affect
significantly the validity of (\ref{slope}). The exact result is expected to
differ from (\ref{slope}) by the numerical factor of the order of $1$ only.
This can be seen by considering the high temperature limit $T\gg 1/\tau $
where one easily obtains the relation between the two-state and three-state
approximation results $S_{3}/S_{2}=2$.

To conclude, we argued in this paper that nonadiabatic effects may limit the
accuracy of the quantization of the current, driven by surface acoustic
waves. The time-dependence can be described effectively by a single
parameter $\tau $, the decay time of the tunneling coupling between 2DEG and
moving minimum of SAW-induced potential. The corresponding energy scale $%
\hbar /\tau $ controls both the slope of the plateaus, and the width of the
transition region between the plateaus. 

\section*{ACKNOWLEDGMENTS}
We are grateful to H. Bruus, Yu. Galperin, L. Glazman, A.-P. Jauho, and A.
Kristensen for very valuable discussions. This work was supported by EC
through the contracts SMT4-CT96-2049 and SMT4-CT98-9030, by the National
Science Foundation under the Grant No. PHY94-07194, and by Welch Foundation.


\begin{thebibliography}{9}
\bibitem{frequency}  For $f\sim 3\;{\rm GHz}$, $N_{0}=1$, Eq.(\ref{quantization}) gives 
$I_{ae}\sim 1\,{\rm nA}$ , while other devices, such as single-electron
pumps, are bound to operate at low frequencies of the order of $1\;{\rm MHz}$, 
and, therefore, deliver quantized currents in the ${\rm pA}$ range.
\bibitem{orthodox}  Time-independent version of this model is widely used in
the theory of the Coulomb blockade (see the review by L. I. Glazman in this
issue and references therein).
\end{thebibliography}
\end{document}